\documentclass[12pt,a4paper]{article}
\usepackage{amsmath}
\usepackage{bm}
\usepackage{amsfonts}
\usepackage{graphicx}
\usepackage[normal]{caption2}
\usepackage{subfigure}
\usepackage{rotating}
\usepackage{citesort}
\usepackage{lscape}
\usepackage{section}
\begin{document}
\renewcommand\arraystretch{1.1}
\setlength{\abovecaptionskip}{0.1cm}
\setlength{\belowcaptionskip}{0.5cm}
\pagestyle{empty}
\newpage
\pagestyle{plain} \setcounter{page}{1} \setcounter{lofdepth}{2}
\begin{center} {\large\bf Role of momentum correlations in the properties of fragments produced in heavy-ion collisions}\\
\vspace*{0.4cm}
{\bf Sakshi Gautam}\footnote{Email:~sakshigautm@gmail.com} {and \bf Rajni Kant}\\
{\it  Department of Physics, Panjab University, Chandigarh -160
014, India.\\}
\end{center}
The role of momentum correlations is studied in the properties of
light and medium mass fragments by imposing momentum cut in
clustering the phase space. Our detailed investigation shows that
momentum cut has major role to play in the properties of
fragments.
\newpage
\baselineskip 20pt
\section{Introduction}
 \par
Intermediate energy heavy-ion collisions produce rich amount of
information on the correlations and fluctuations and eventually on
the dynamics and interactions among the nucleons. The breaking of
nuclei, i.e., multifragmentation, is one of the rare phenomena
that has attracted major attention in recent years \cite{bege}.
The physics behind multifragmentation is so complicated that many
different theoretical approaches have been developed
\cite{bege,aich1,qmd1,dorso}. Since no theoretical model simulates
fragments, one needs afterburners to identify clusters. Since
correlations and fluctuations are the main features of the
molecular dynamics model, the quantum molecular dynamics (QMD)
model is very successful in explaining the phenomena of
multifragmentation. Once the phase space is accessible, one
generally clusterize the phase space with simple spatial
correlation method where one binds the nucleons in a fragment they
lie within a distance of 4 fm. This method is known as minimum
spanning tree (MST) method \cite{jsingh}. At the same time
fragments formed in MST method will be highly unstable (especially
in central collisions) as there the two nucleons may not be well
formed and therefore can be unstable that will decay after a
while. In order to filter out such unstable fragments, we impose
another cut in terms of relative momentum of nucleons. This
method, dubbed as minimum spanning tree with momentum cut (MSTP)
method was discussed by Puri \emph{et al.} \cite{kumar1}. In our
recent work, we study the role of momentum cut on fragment
structure \cite{rajni}. We also studied the role of colliding
geometry on the fragmentation when momentum cut is being imposed.
No study exits in literature to see the role of momentum cut on
the various fragment properties like the rapidity distribution,
p$_{t}$ spectra and E$_{rat}$. So the the present paper, we plan
to see the role of momentum cut on various fragment properties and
to investigate how these properties vary with impact parameter.
The present study is carried out within the framework of QMD model
\cite{aich1,qmd1} which is described in the following section.

\section{The Formalism}
\subsection{Quantum Molecular dynamics (QMD) model}
\par
We describe the time evolution of a heavy-ion reaction within the
framework of Quantum Molecular Dynamics (QMD) model
\cite{aich1,qmd1} which is based on a molecular dynamics picture.
This model has been successful in explaining collective flow
\cite{sood2}, elliptic flow \cite{kumar3}, multifragmentation
\cite{dhawan} as well as dense and hot matter \cite{fuchs}. Here
each nucleon is represented by a coherent state of the form
\begin{equation}
\phi_{\alpha}(x_1,t)=\left({\frac {2}{L \pi}}\right)^{\frac
{3}{4}} e^{-(x_1-x_{\alpha }(t))^2}
e^{ip_{\alpha}(x_1-x_{\alpha})} e^{-\frac {i p_{\alpha}^2 t}{2m}}.
\label {e1}
\end{equation}
Thus, the wave function has two time dependent parameters
$x_{\alpha}$ and $p_{\alpha}$.  The total n-body wave function is
assumed to be a direct product of coherent states:
\begin{equation}
\phi=\phi_{\alpha}
(x_1,x_{\alpha},p_{\alpha},t)\phi_{\beta}(x_2,x_{\beta},
p_{\beta},t)....,         \label {e2}
\end{equation}
where antisymmetrization is neglected. One should, however, keep
in the mind that the Pauli principle, which is very important at
low incident energies, has been taken into account. The initial
values of the parameters are chosen in a way that the ensemble
($A_T$+$A_P$) nucleons give a proper density distribution as well
as a proper momentum distribution of the projectile and target
nuclei. The time evolution of the system is calculated using the
generalized variational principle. We start out from the action
\begin{equation}
S=\int_{t_1}^{t_2} {\cal {L}} [\phi,\phi^{*}] d\tau, \label {e3}
\end{equation}
with the Lagrange functional
\begin{equation}
{\cal {L}} =\left(\phi\left|i\hbar \frac
{d}{dt}-H\right|\phi\right), \label {e4}
\end{equation}
where the total time derivative includes the derivatives with
respect to the parameters. The time evolution is obtained by the
requirement that the action is stationary under the allowed
variation of the wave function
\begin{equation}
\delta S=\delta \int_{t_1}^{t_2} {\cal {L}} [\phi ,\phi^{*}] dt=0.
\label{e5}
\end{equation}
If the true solution of the Schr\"odinger equation is contained in
the restricted set of wave function
$\phi_{\alpha}\left({x_{1},x_{\alpha},p_{\alpha}}\right),$ this
variation of the action gives the exact solution of the
Schr\"odinger equation. If the parameter space is too restricted,
we obtain that wave function in the restricted parameter space
which comes close to the solution of the Schr\"odinger equation.
Performing the variation with the test wave function (2), we
obtain for each parameter $\lambda$ an Euler-Lagrange equation;
\begin{equation}
\frac{d}{dt} \frac{\partial {\cal {L}}}{\partial {\dot
{\lambda}}}-\frac{\partial \cal {L}} {\partial \lambda}=0.
\label{e6}
\end{equation}
For each coherent state and a Hamiltonian of the form, \\

$H=\sum_{\alpha}
\left[T_{\alpha}+{\frac{1}{2}}\sum_{\alpha\beta}V_{\alpha\beta}\right]$,
the Lagrangian and the Euler-Lagrange function can be easily
calculated
\begin{equation}
{\cal {L}} = \sum_{\alpha}{\dot {\bf x}_{\alpha}} {\bf
p}_{\alpha}-\sum_{\beta} \langle{V_{\alpha
\beta}}\rangle-\frac{3}{2Lm}, \label{e7}
\end{equation}
\begin{equation}
{\dot {\bf x}_{\alpha}}=\frac{{\bf
p}_\alpha}{m}+\nabla_{p_{\alpha}}\sum_{\beta} \langle{V_{\alpha
\beta}}\rangle, \label {e8}
\end{equation}
\begin{equation}
{\dot {\bf p}_{\alpha}}=-\nabla_{{\bf x}_{\alpha}}\sum_{\beta}
\langle{V_{\alpha \beta}}\rangle. \label {e9}
\end{equation}
Thus, the variational approach has reduced the n-body
Schr\"odinger equation to a set of 6n-different equations for the
parameters which can be solved numerically. If one inspects  the
formalism carefully, one finds that the interaction potential
which is actually the Br\"{u}ckner G-matrix can be divided into
two parts: (i) a real part and (ii) an imaginary part. The real
part of the potential acts like a potential whereas imaginary part
is proportional to the cross section.

In the present model, interaction potential comprises of the
following terms:
\begin{equation}
V_{\alpha\beta} = V_{loc}^{2} + V_{loc}^{3} + V_{Coul} + V_{Yuk}
 \label {e10}
\end {equation}
$V_{loc}$ is the Skyrme force whereas $V_{Coul}$, $V_{Yuk}$ and
$V_{MDI}$ define, respectively, the Coulomb, and Yukawa
potentials. The Yukawa term separates the surface which also plays
the role in low energy processes like fusion and cluster
radioactivity \cite{puri}. The expectation value of these
potentials is calculated as
\begin{eqnarray}
V^2_{loc}& =& \int f_{\alpha} ({\bf p}_{\alpha}, {\bf r}_{\alpha},
t) f_{\beta}({\bf p}_{\beta}, {\bf r}_{\beta}, t)V_I ^{(2)}({\bf
r}_{\alpha}, {\bf r}_{\beta})
\nonumber\\
&  & \times {d^{3} {\bf r}_{\alpha} d^{3} {\bf r}_{\beta}
d^{3}{\bf p}_{\alpha}  d^{3}{\bf p}_{\beta},}
\end{eqnarray}
\begin{eqnarray}
V^3_{loc}& =& \int  f_{\alpha} ({\bf p}_{\alpha}, {\bf
r}_{\alpha}, t) f_{\beta}({\bf p}_{\beta}, {\bf r}_{\beta},t)
f_{\gamma} ({\bf p}_{\gamma}, {\bf r}_{\gamma}, t)
\nonumber\\
&  & \times  V_I^{(3)} ({\bf r}_{\alpha},{\bf r}_{\beta},{\bf
r}_{\gamma}) d^{3} {\bf r}_{\alpha} d^{3} {\bf r}_{\beta} d^{3}
{\bf r}_{\gamma}
\nonumber\\
&  & \times d^{3} {\bf p}_{\alpha}d^{3} {\bf p}_{\beta} d^{3} {\bf
p}_{\gamma}.
\end{eqnarray}
where $f_{\alpha}({\bf p}_{\alpha}, {\bf r}_{\alpha}, t)$ is the
Wigner density which corresponds to the wave functions (eq. 2). If
we deal with the local Skyrme force only, we get
{\begin{equation} V^{Skyrme} = \sum_{{\alpha}=1}^{A_T+A_P}
\left[\frac {A}{2} \sum_{{\beta}=1} \left(\frac
{\tilde{\rho}_{\alpha \beta}}{\rho_0}\right) + \frac
{B}{C+1}\sum_{{\beta}\ne {\alpha}} \left(\frac {\tilde
{\rho}_{\alpha \beta}} {\rho_0}\right)^C\right].
\end{equation}}

Here A, B and C are the Skyrme parameters which are defined
according to the ground state properties of a nucleus. Different
values of C lead to different equations of state. A larger value
of C (= 380 MeV) is often dubbed as stiff equation of state.The
finite range Yukawa ($V_{Yuk}$) and effective Coulomb potential
($V_{Coul}$) read as:
\begin{equation}
V_{Yuk} = \sum_{j, i\neq j} t_{3}
\frac{exp\{-|\textbf{r}_{\textbf{i}}-\textbf{r}_{\textbf{j}}|\}/\mu}{|\textbf{r}_{\textbf{i}}-\textbf{r}_{\textbf{j}}|/\mu},
\end{equation}
\begin{equation}
V_{Coul} = \sum_{j, i\neq
j}\frac{Z_{eff}^{2}e^{2}}{|\textbf{r}_{\textbf{i}}-\textbf{r}_{\textbf{j}}|}.
\end{equation}
\par
The Yukawa interaction (with $t_{3}$= -6.66 MeV and $\mu$ = 1.5
fm) is essential for the surface effects. The relativistic effect
does not play role in low incident energy of present interest
\cite{lehm}.
\par
The phase space of nucleons is stored at several time steps. The
QMD model does not give any information about the fragments
observed at the final stage of the reaction. In order to construct
 the fragments, one needs
clusterization algorithms. We shall concentrate here on the MST
and MSTP methods.
\par
 According to MST method
\cite{jsingh}, two nucleons are allowed to share the same fragment
if their centroids are closer than a distance $r_{min}$,
\begin{equation}
|\textbf{r}_{\textbf{i}}-\textbf{r}_{\textbf{j}}| \leq r_{min}.
\end{equation}
where $\textbf{r}_{\textbf{i}}$ and $\textbf{r}_{\textbf{j}}$ are
the spatial positions of both nucleons and r$_{min}$ taken to be
4fm.
\par
 For MSTP method,we impose a additional cut in the
momentum space, i.e., we allow only those nucleons to form a
fragment which in addition to equation(16) also satisfy
\begin{eqnarray}
|\textbf{p}_{\textbf{i}}-\textbf{p}_{\textbf{j}}| \leq p_{min},
\end{eqnarray}
where p$_{min}$ = 150 MeV/c.
\par
\section{Results and Discussion}
We simulated the reactions of $^{12}$C+$^{12}$C and
$^{40}$Ca+$^{40}$Ca  at 100 MeV/nucleon at central and peripheral
colliding geometries, i.e., at  $\hat{b}$ = 0.0 and 0.8,
respectively. We use a soft equation of state with standard
energy-dependent Cugon cross section.
\par

\begin{figure}[!t]
\centering
 \vskip 1cm
\includegraphics[angle=0,width=12cm]{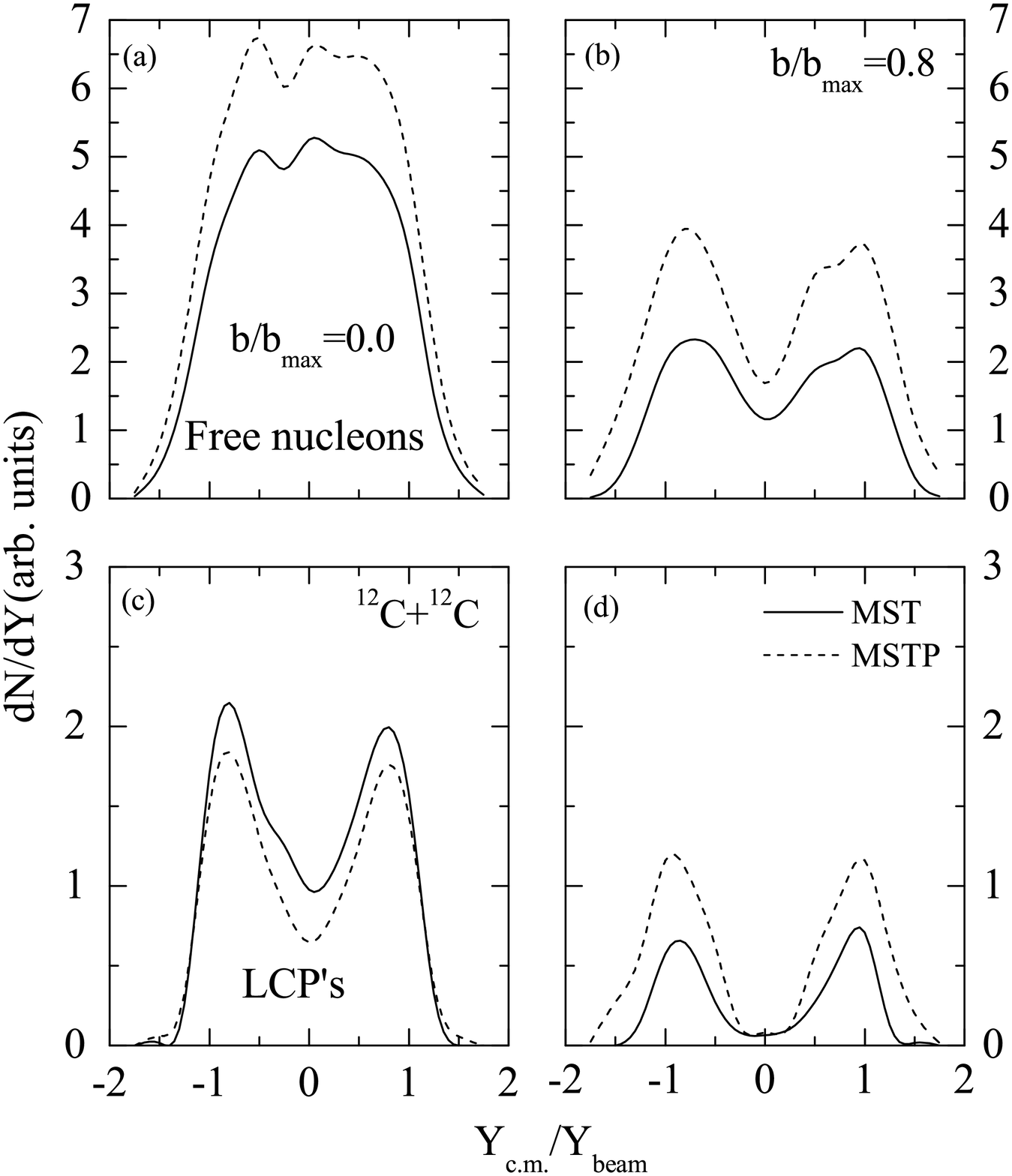}
 \vskip -0cm \caption{ The rapidity distribution for free nucleons and LCPs for the reaction of $^{12}$C+$^{12}$C at incident energy of
 100 MeV/nucleon at central (left panel) and peripheral (right) geometries with MST and MSTP methods. }\label{fig1}
\end{figure}

\begin{figure}[!t]
\centering \vskip 1cm
\includegraphics[angle=0,width=12cm]{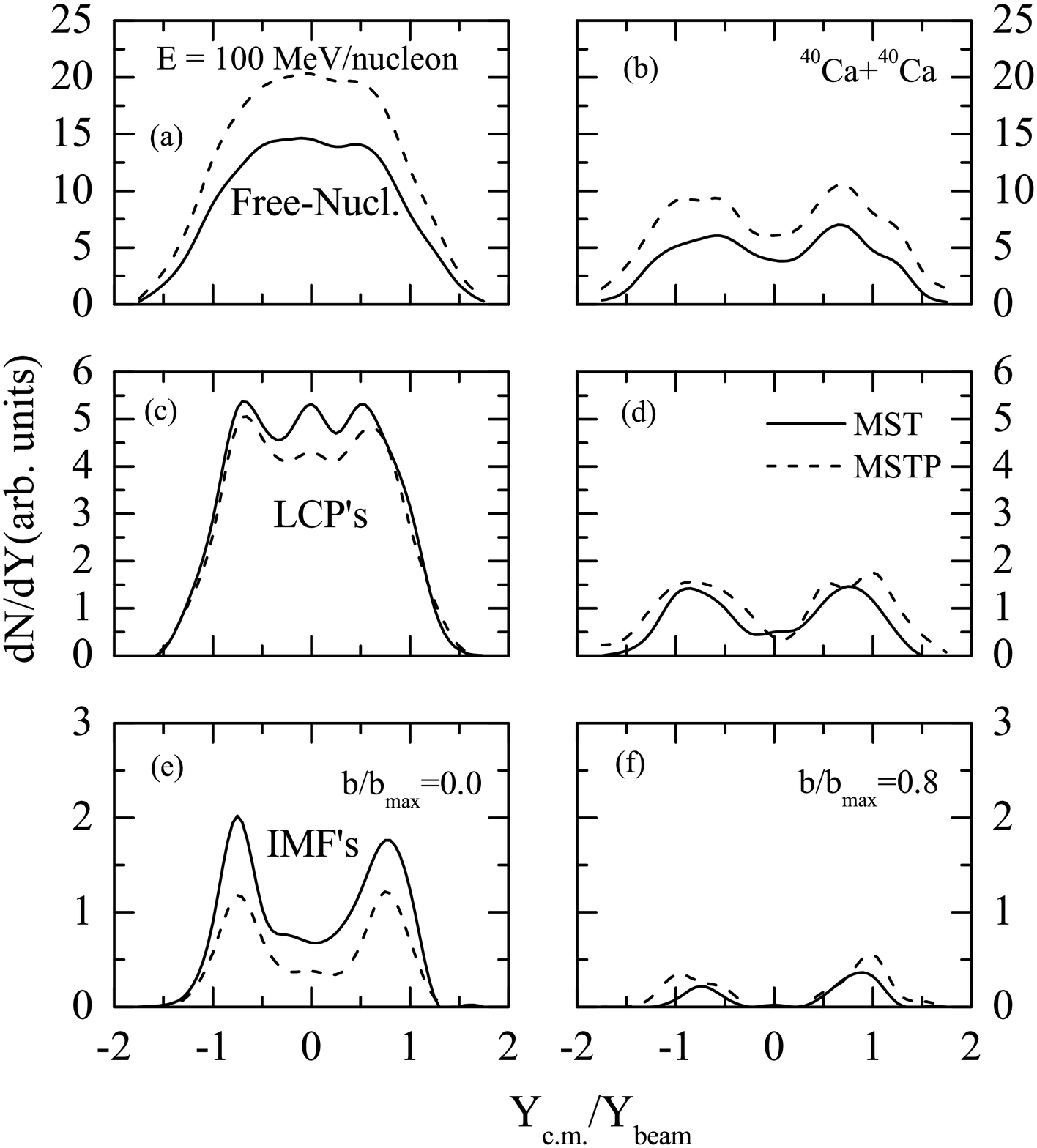}
\vskip -0cm \caption{Same as Fig. 1 but for the reaction of
$^{40}$Ca+$^{40}$Ca.}\label{fig2}
\end{figure}

 \begin{figure}[!t]
\centering \vskip 1cm
\includegraphics[angle=0,width=12cm]{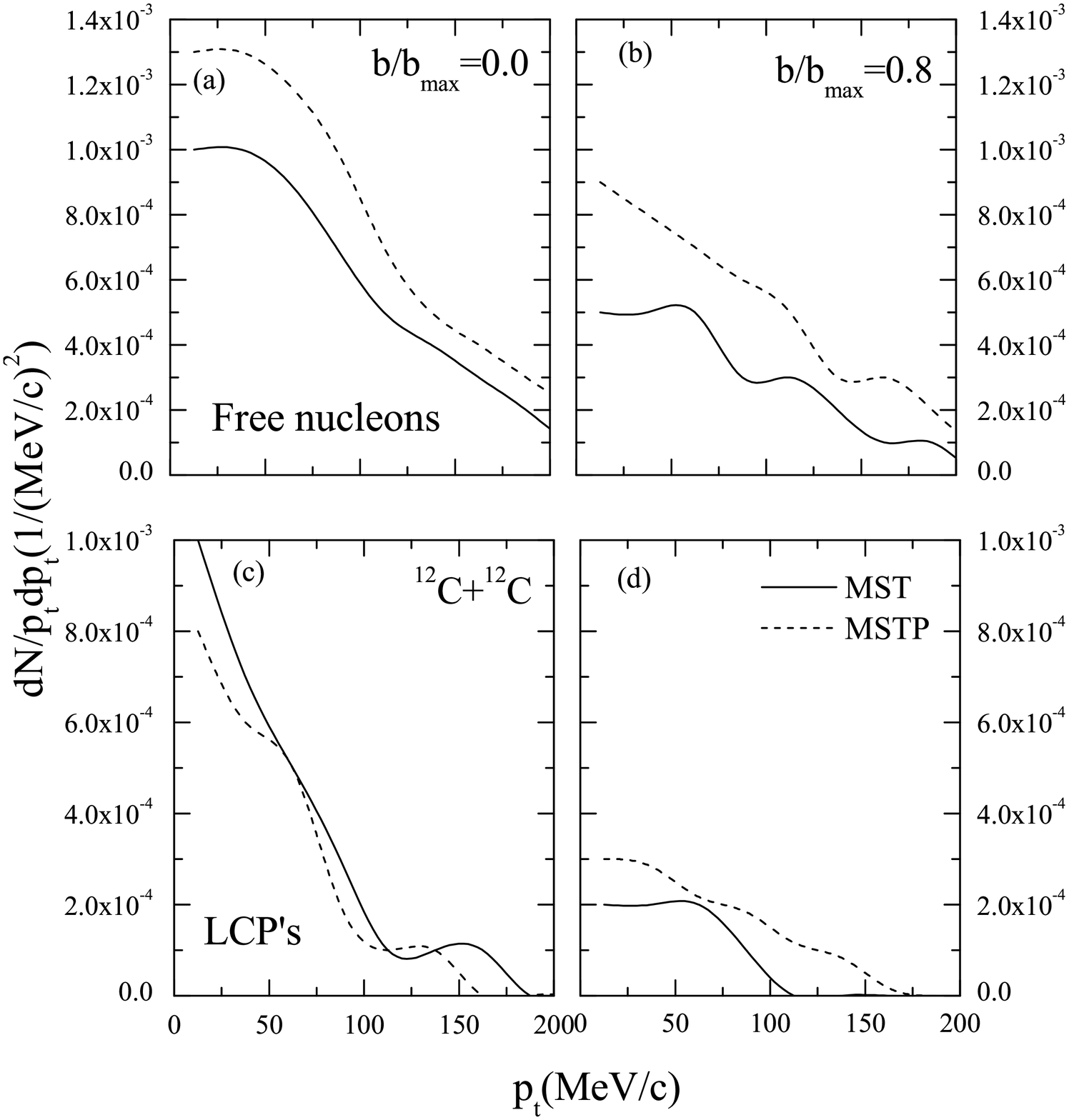}
\vskip -0cm \caption{The $\frac{dN}{p_{t}dp_{t}}$
(1/(MeV/c)$^{2}$) as a function of transverse energy p$_{t}$ for
the free nucleons and LCPs for the reaction of $^{12}$C+$^{12}$C
at central (left panel) and peripheral (right) geometries with MST
and MSTP methods. Lines have same meaning as in Fig.
1.}\label{fig3}
\end{figure}

\begin{figure}[!t] \centering
 \vskip 1cm
\includegraphics[angle=0,width=12cm]{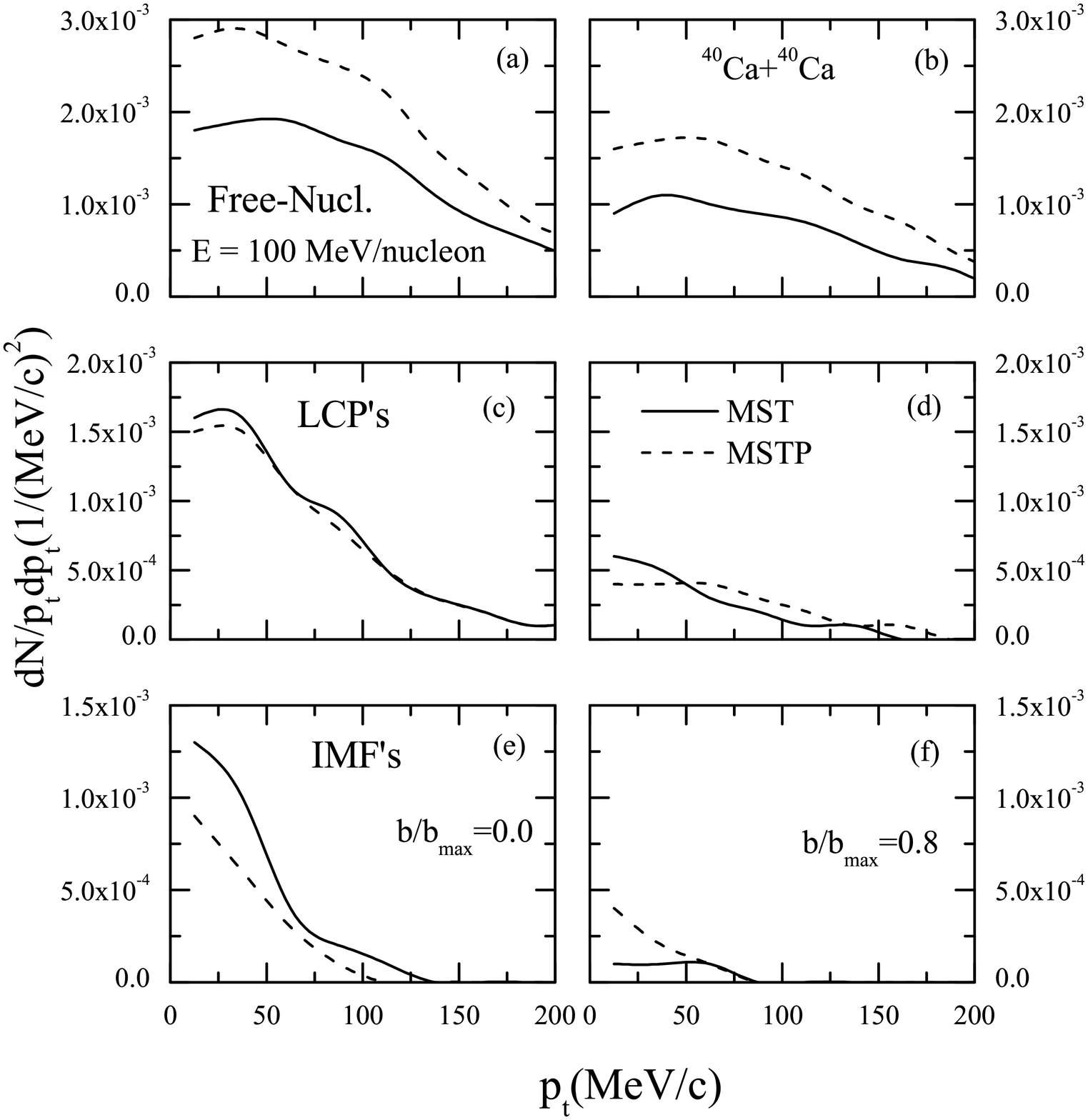}
 \vskip -0cm \caption{ Same as Fig. 3 but for the reaction of
$^{40}$Ca+$^{40}$Ca.}\label{fig5}
\end{figure}

\begin{figure}[!t]
\centering
 \vskip 1cm
\includegraphics[angle=0,width=12cm]{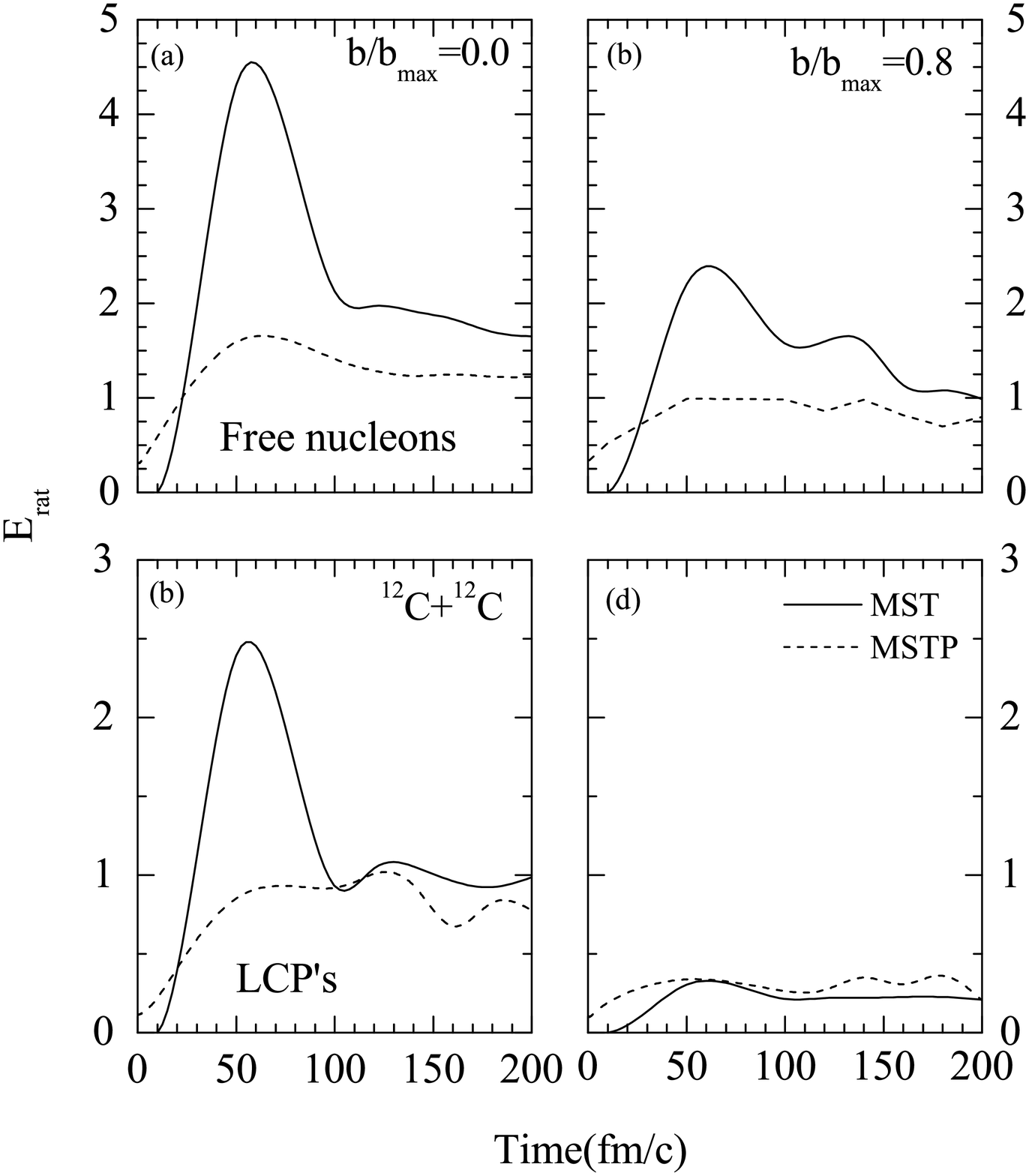}
 \vskip -0cm \caption{ The time evolution of E$_{rat}$ for free nucleons and LCPs for the reaction of
$^{12}$C+$^{12}$C.}\label{fig6}
\end{figure}

\begin{figure}[!t]
\centering
 \vskip 1cm
\includegraphics[angle=0,width=12cm]{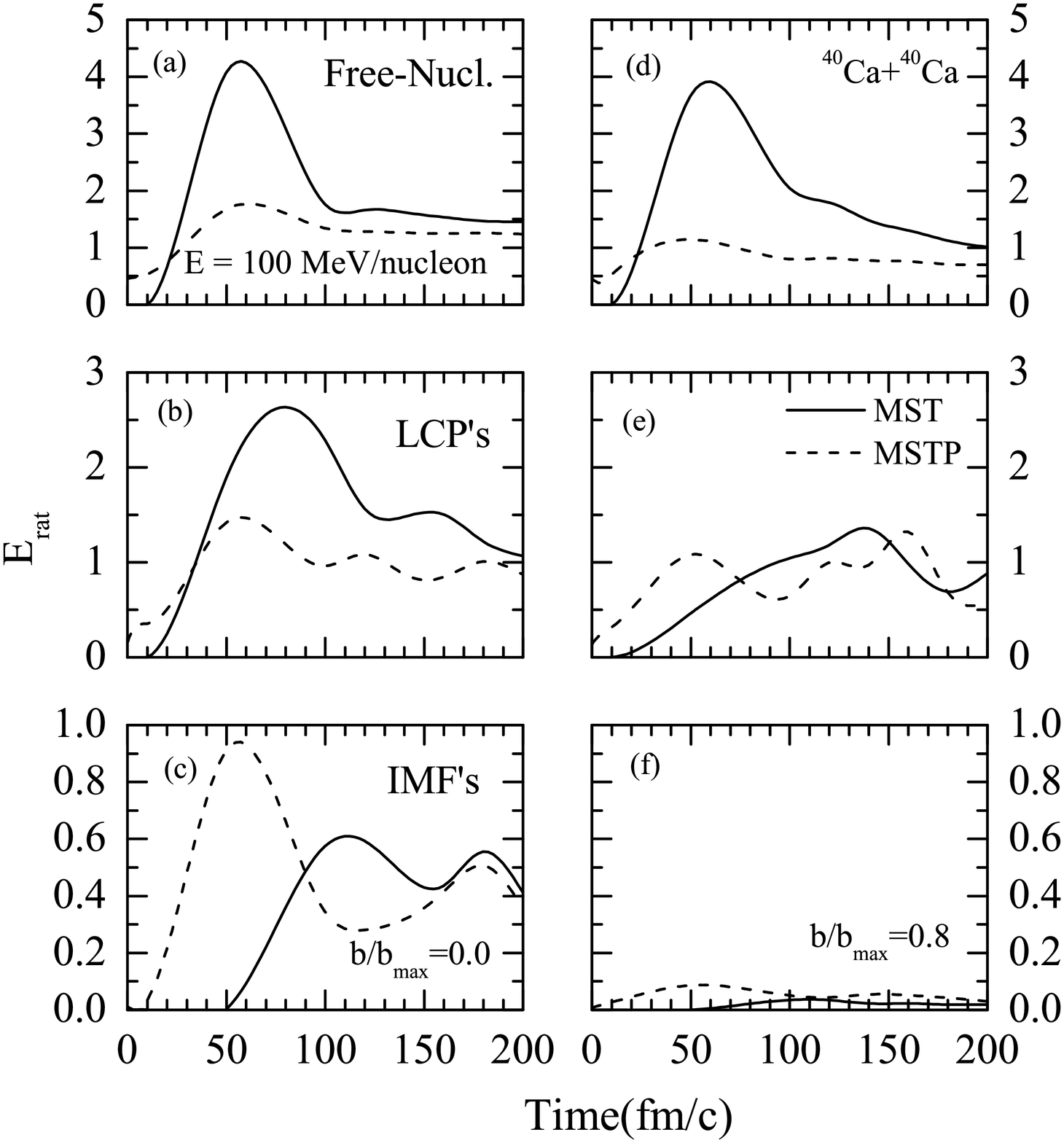}
 \vskip -0cm \caption{ Same as Fig. 5 but for the reaction of
$^{40}$Ca+$^{40}$Ca.}\label{fig6}
\end{figure}

\par
In figure 1, we display the rapidity distribution of free nucleon
and LCP's for the reaction of $^{12}$C+$^{12}$C at energy of 100
MeV at central (left panel) and peripheral (right) colliding
geometry. The solid and dashed lines indicate the calculations of
MST and MSTP methods, respectively. From the figure, we see that
there is quantitative difference in the results of MST and MSTP
methods, though qualitatively, both methods give similar behaviour
of the rapidity distribution of nucleons and fragments.
\par
For central collisions (left panel), we see that the peak of dN/dY
plot is pronounced for the MSTP method, thus, indicating enhanced
production of free nucleons in MSTP method as compared to MST
method. This is due to the fact that in the MST method, we have a
single big fragment because of no restriction is being imposed on
the relative momentum of nucleons forming fragments. The
production of LCP's is more with MSTP method compared to MST
method which is supported by Ref. \cite{kumar1,rajni}. At
peripheral collisions, the behaviour of the rapidity plots of free
nucleons is similar as for the central one whereas the trend
reverses for the LCP's plot. For peripheral collisions, we have
greater production with MSTP method.
\par
In figure 2, we display the rapidity distribution of free
nucleons, LCP's and IMF's for the reaction of $^{40}$Ca+$^{40}$Ca.
Left(right) panels display the results for b/b$_{max}$=0.0 (0.8).
We find similar behaviour of free nucleons and LCP's as reported
for the reaction of $^{12}$C+$^{12}$C. The IMF's also follow the
similar trend as for LCP's, i.e., we have more (less) production
of IMF's with MST method at central (peripheral) collisions.
\par
In figures 3 and 4, we display dN/p$_{t}$dp$_{t}$ versus p$_{t}$
for the reaction of $^{12}$C+$^{12}$C and $^{40}$Ca+$^{40}$Ca,
respectively. We see that dN/p$_{t}$dp$_{t}$ spectra follow the
similar behaviour for both MST and MSTP methods. We have a higher
peak in the spectra of free nucleons with MST method at both the
colliding geometries. The difference between MST and MSTP methods
in spectra of LCP's is less significant. Similar behaviour is also
observed for the reaction of $^{40}$Ca+$^{40}$Ca.
\par
In figure 5, we display the time evolution of E$_{rat}$ of free
nucleons and LCP's for the reaction of $^{12}$C+$^{12}$C at
central (left panel) and peripheral (right) colliding geometry.
For MST and MSTP methods, we find a significant difference between
MST and MSTP methods for both free nucleons and LCP's. The
difference is more for the central collisions as compared to the
peripheral one.
\par
In figure 6, we display time evolution of E$_{rat}$ of free
nucleons, LCP's and IMF's for the reaction of $^{40}$Ca+$^{40}$Ca
at central (left panel) and peripheral(right) collisions. The
solid (dashed) lines represent the results of MST (MSTP) method.
From figure, we find a significant difference of E$_{rat}$ with
MST and MSTP method as in case of the reaction of
$^{12}$C+$^{12}$C. We also find that the difference between MST
and MSTP reduces at peripheral colliding geometries.

\section{Summary}
 Using the quantum molecular dynamic model, we studied the role of
 momentum correlations in the properties of fragments. This was achieved by
 imposing cut in momentum space during the process of clusterization. We find that this cut yields significant
 difference in the fragment properties of system at all colliding
 geometries.

\section{Acknowledgement}
This work is done under the supervision of Dr. Rajeev K. Puri,
Department of Physics, Panjab University, Chandigarh, India. This
work has been supported by a grant from Centre of Scientific and
Industrial Research (CSIR), Govt. of India.

\end{document}